\documentclass[12pt,thmsa,titlepage]{article}
\usepackage[dvips]{graphicx}
\usepackage{amsmath}
\usepackage{amsfonts}
\usepackage{amssymb}

\begin{document}

\author{S. Gov\thanks{Also with the Center for Technological Education Holon, 52
Golomb St., P.O.B 305, Holon 58102, Israel.} and S. Shtrikman\thanks{Also with
the Department of Physics, University of California, San Diego, La Jolla,
92093 CA, USA.}\\The Department of Electronics, \\Weizmann Institute of Science,\\Rehovot 76100, Israel}
\title{Dynamic Stability of The Time-averaged Orbiting Potential Trap: Exact
Classical Analysis.}
\maketitle
\begin{abstract}
We calculate \emph{exactly} the modes of motion of the Time-averaged Orbiting
Potential (TOP) trap with its four degrees of freedom, namely the three
translations and the spin, taken into account. We find that, when gravity is
neglected, there are \emph{two} parameters in the problem namely, the angular
velocity of the rotating field and its strength. We present the stability
diagram in these parameters. We find the mode frequencies calculated from the
time-averaged potential model used by the inventors of the TOP is an excellent
approximation to our exact results. However, for other parameters, this may
not be the case.
\end{abstract}

\section{\bigskip Introduction.}

The first observation of Bose-Einstein condensate (BEC) \cite{first} was done
in the so-called Time-averaged Orbiting Potential (TOP) trap. Although later
other magnetic traps, optical plug, and even an all-optical one\cite{allopt}
were used, the ingenious TOP trap continues to be a workhorse in the
trade\cite{han} because of its unique advantage as pointed out by
\cite{cornell}. It seemed to us thus worthwhile to extend the analysis given
in Ref.\cite{cornell}, which is based on the concept of a time-averaged
adiabatic potential, by studying the \emph{exact} motion of all the degrees of
freedom in the problem, namely the three translational degrees of freedom and
the spin degree of freedom of the particle\cite{deluca}. This has the
advantage of allowing the calculation of limitations required by stability. We
find that, when gravity $g$ is neglected, there are \emph{two} parameters in
the problem, the angular velocity of the rotating field $\Omega$ and its
strength $\alpha$, and we present the stability diagram in the $\alpha
$-$\Omega$ plane. The parameters reported in Ref.\cite{cornell} are shown to
be well inside the stable region. Their calculated frequencies agree
excellently with our exact frequencies for their experiment. When the limit of
high field strength is taken, our analytic results reduce to their formulae.
The stability diagram that we found shows that the TOP trap is very flexible
for the experimentalist in terms of allowed parameters $\alpha$ and $\Omega$.
Our treatment is classical, but we also discuss what is involved in a
quantum-mechanical calculation.

The structure of this paper is as follows: In Sec.\ref{sec2} we first define
the problem, write down the equations of motion and find stationary solutions
of these equations. Next, we perturb the stationary solutions, linearize the
equations of motion and derive the secular equation which determines the
frequencies of the various possible modes. The secular equation is given as a
function of $\alpha$, $\Omega$ and $g$. In Sec.\ref{sec3} we use the secular
equation to find the mode frequencies for the TOP trap described in
Ref.\cite{cornell}, and compare our exact results with their calculations and
measurements. In Sec.\ref{sec4} we specify to the case $g=0$ and derive
approximate expressions for the mode frequencies in the limit where $\alpha$
is large, starting from the secular equation. These are found to agree with
the corresponding expressions given in Ref.\cite{cornell}. \ In Sec.\ref{sec5}
we present the stability region in the $\alpha$-$\Omega$ plane (for $g=0$) for
which stable trapping occurs, and comment on its implications and limitations.
Finally, in Sec.\ref{sec6} we summarize our results and discuss briefly the
related quantum-mechanical problem.

\section{Derivation of secular equation.\label{sec2}}

\subsection{Mathematical formulation and physical
interpretation.\label{sec2.1}}

We consider a particle of mass $m$, intrinsic spin $\mathbf{S=}S\mathbf{\hat
{n}}$ and magnetic moment $\mathbf{\mu=-}\mu\mathbf{\hat{n}}$, moving in 3D
space in the presence of uniform gravitation $-G\mathbf{\hat{z}}$ and an
inhomogeneous time-dependent magnetic field given by%
\begin{equation}
\mathbf{H}=H^{\prime}\left(  \mathbf{-}\frac{1}{2}\rho\mathbf{\hat{\rho}%
}+h\mathbf{\hat{z}}\right)  \mathbf{+}H\left(  \cos\left(  \varphi-\Omega
_{r}t\right)  \mathbf{\hat{\rho}}-\sin\left(  \varphi-\Omega_{r}t\right)
\mathbf{\hat{\varphi}}\right)  . \label{d0}%
\end{equation}
Here, $\left(  h,\rho,\varphi\right)  $ are the height, radial distance and
polar angle of the center of mass of the particle in cylindrical coordinate
system, $\left(  \mathbf{\hat{z},\hat{\rho}},\mathbf{\hat{\varphi}}\right)  $
are the corresponding unit vectors, $H$ is the strength of the (uniform)
time-dependent field rotating at an angular velocity $\Omega_{r}$ about the
$z$-axis, and $H^{\prime}$ is the time-independent field gradient in the $z$-direction.

The equations of motion for the center of mass of the particle $\mathbf{r}%
=z\mathbf{\hat{z}+}\rho\mathbf{\hat{\rho}}$ are%
\begin{equation}
m\dfrac{d^{2}\mathbf{r}}{dt^{2}}=-\mu\mathbf{\nabla}\left(  \mathbf{\hat
{n}\cdot H}\right)  -mG\mathbf{\hat{z}}, \label{d1}%
\end{equation}
and the evolution of its spin is determined by%
\begin{equation}
S\dfrac{d\mathbf{\hat{n}}}{dt}=-\mu\mathbf{\hat{n}\times H.} \label{d2}%
\end{equation}
In the following, it is convenient to express $\mathbf{\hat{n}}$- a unit
vector in the direction of the spin, in terms of its components along the
$\left(  \mathbf{\hat{z},\hat{\rho}},\mathbf{\hat{\varphi}}\right)  $
directions. We denote these by $n_{z}$, $n_{\rho}$ and $n_{\varphi}$,
respectively. Note however, that the unit vectors $\left(  \mathbf{\hat{\rho}%
},\mathbf{\hat{\varphi}}\right)  $ themselves depend on time according to%
\begin{align}
\dfrac{d\mathbf{\hat{\rho}}}{dt}  &  =\dfrac{d\varphi}{dt}\mathbf{\hat
{\varphi},}\label{d3}\\
\dfrac{d\mathbf{\hat{\varphi}}}{dt}  &  =-\dfrac{d\varphi}{dt}\mathbf{\hat
{\rho}}\text{.}\nonumber
\end{align}
Substitution of Eq.(\ref{d0}) into Eqs.(\ref{d1}) and (\ref{d2}), and making
use of Eq.(\ref{d3}) yields%
\begin{equation}%
\begin{array}
[c]{c}%
\dfrac{d^{2}\rho}{dt^{2}}-\rho\left(  \dfrac{d\varphi}{dt}\right)  ^{2}%
=\dfrac{\mu H^{\prime}}{2m}n_{\rho}\\
2\dfrac{d\rho}{dt}\dfrac{d\varphi}{dt}+\rho\dfrac{d^{2}\varphi}{dt^{2}}%
=\dfrac{\mu H^{\prime}}{2m}n_{\varphi}\\
\dfrac{d^{2}h}{dt^{2}}=-\dfrac{\mu H^{\prime}}{m}n_{z}-G\\
\dfrac{dn_{\rho}}{dt}-\dfrac{d\varphi}{dt}n_{\varphi}=-\dfrac{\mu}{S}\left[
H^{\prime}hn_{\varphi}+Hn_{z}\sin\left(  \varphi-\Omega_{r}t\right)  \right]
\\
\dfrac{dn_{\varphi}}{dt}+\dfrac{d\varphi}{dt}n_{\rho}=-\dfrac{\mu}{S}\left[
Hn_{z}\cos\left(  \varphi-\Omega_{r}t\right)  -\dfrac{1}{2}\rho n_{z}%
H^{\prime}-n_{\rho}H^{\prime}h\right] \\
\dfrac{dn_{z}}{dt}=-\dfrac{\mu}{S}\left[  -n_{\varphi}H\cos\left(
\varphi-\Omega_{r}t\right)  +\frac{1}{2}\rho n_{\varphi}H^{\prime}-n_{\rho
}H\sin\left(  \varphi-\Omega_{r}t\right)  \right]  .
\end{array}
\label{d4}%
\end{equation}
Eqs.(\ref{d4}) are almost the equations of motion of the particle in a
coordinate system which is rotating with the field. The only difference is in
the definition of the angle $\varphi$, which is measured with respect to the
\emph{fixed} $\mathbf{\hat{x}}$ axis rather than with respect to the axis
defined by the rotating field. To show this we rewrite the equations of motion
in the rotating frame by substituting%
\begin{equation}%
\begin{array}
[c]{c}%
\mathbf{r}\rightarrow\mathbf{r,}\\
\dfrac{d\mathbf{r}}{dt}\rightarrow\dfrac{d\mathbf{r}}{dt}\mathbf{+}\Omega
_{r}\mathbf{\hat{z}}\times\mathbf{r,}\\
\dfrac{d^{2}\mathbf{r}}{dt^{2}}\rightarrow\dfrac{d^{2}\mathbf{r}}{dt^{2}%
}+2\Omega_{r}\mathbf{\hat{z}}\times\dfrac{d\mathbf{r}}{dt}+\Omega_{r}%
^{2}\mathbf{\hat{z}\times}\left(  \mathbf{\hat{z}}\times\mathbf{r}\right)  ,\\
\left(  \dfrac{d\mathbf{\hat{n}}}{dt}\right)  \rightarrow\left(
\dfrac{d\mathbf{\hat{n}}}{dt}\right)  +\Omega_{r}\mathbf{\hat{z}}%
\times\mathbf{\hat{n}.}%
\end{array}
\label{d5}%
\end{equation}
This brings the equations of motion Eq.(\ref{d1}) and (\ref{d2}) into%
\begin{equation}
m\left[  \dfrac{d^{2}\mathbf{r}}{dt^{2}}+2\Omega_{r}\mathbf{\hat{z}}%
\times\dfrac{d\mathbf{r}}{dt}+\Omega_{r}^{2}\mathbf{\hat{z}\times}\left(
\mathbf{\hat{z}}\times\mathbf{r}\right)  \right]  =-\mu\mathbf{\nabla}\left[
\mathbf{\hat{n}\cdot}\left(  \mathbf{H-}\dfrac{S}{\mu}\Omega_{r}%
\mathbf{\hat{z}}\right)  \right]  -mG\mathbf{\hat{z}} \label{d6.0}%
\end{equation}
and%
\begin{equation}
S\dfrac{d\mathbf{\hat{n}}}{dt}=-\mu\mathbf{\hat{n}\times}\left[
\mathbf{H-}\dfrac{S}{\mu}\Omega_{r}\mathbf{\hat{z}}\right]  , \label{d6.2}%
\end{equation}
where now the magnetic field $\mathbf{H}$ is \emph{time-independent} and is
given by%
\begin{equation}
\mathbf{H=}H^{\prime}\left(  \mathbf{-}\frac{1}{2}\rho\mathbf{\hat{\rho}%
}+h\mathbf{\hat{z}}\right)  \mathbf{+}H\mathbf{\hat{x}.} \label{d7}%
\end{equation}
Eqs.(\ref{d6.0}) and (\ref{d6.2}) indicate that in the comoving frame, the
particle is acted upon by an additional uniform magnetic field $-\left(
S\Omega_{r}/\mu\right)  \mathbf{\hat{z}}$, a centrifugal force $-m\Omega
_{r}^{2}\mathbf{\hat{z}\times}\left(  \mathbf{\hat{z}}\times\mathbf{r}\right)
$ and a velocity-dependent Coriolis force $m\mathbf{v}\times\left(
2\Omega_{r}\mathbf{\hat{z}}\right)  $. The action of the Coriolis force may
also be interpreted as the Lorentz force of the inertial field $\sim\Omega
_{r}\mathbf{\hat{z}}$ which acts on the mass of the particle. Finally,
rewriting Eqs.(\ref{d6.0}) and (\ref{d6.2}) in polar coordinates yields
Eqs.(\ref{d4}) with $\varphi-\Omega_{r}t$ replaced by $\varphi$.

As the number of parameters in the problem is relatively large, we rewrite the
equations of motion in terms of normalized coordinates. We thus define%
\[
R_{0}\equiv\left(  \dfrac{S^{2}}{\mu mH^{\prime}}\right)  ^{1/3}%
\]
as the characteristic length-scale in the problem, and%
\[
\Omega_{0}\equiv\left(  \dfrac{\left(  \mu H^{\prime}\right)  ^{2}}%
{mS}\right)  ^{1/3}%
\]
as the characteristic angular velocity. This allows to define the
dimensionless quantities%
\[%
\begin{array}
[c]{c}%
r\equiv\rho/R_{0}\\
z\equiv h/R_{0}\\
\tau\equiv\Omega_{0}t\\
g\equiv G/(\Omega_{0}^{2}R_{0})\\
\Omega\equiv\Omega_{r}/\Omega_{0}\\
\alpha\equiv\mu H/S\Omega_{0},
\end{array}
\]
with which Eqs.(\ref{d4}) become%
\begin{equation}%
\begin{array}
[c]{c}%
\dfrac{d^{2}r}{d\tau^{2}}-r\left(  \dfrac{d\varphi}{d\tau}\right)  ^{2}%
=\dfrac{1}{2}n_{\rho}\\
2\dfrac{dr}{d\tau}\dfrac{d\varphi}{d\tau}+r\dfrac{d^{2}\varphi}{d\tau^{2}%
}=\dfrac{1}{2}n_{\varphi}\\
\dfrac{d^{2}z}{d\tau^{2}}=-n_{z}-g\\
\dfrac{dn_{\rho}}{d\tau}-\dfrac{d\varphi}{d\tau}n_{\varphi}=-zn_{\varphi
}-\alpha n_{z}\sin\left(  \varphi-\Omega\tau\right) \\
\dfrac{dn_{\varphi}}{d\tau}+\dfrac{d\varphi}{d\tau}n_{\rho}=-\alpha n_{z}%
\cos\left(  \varphi-\Omega\tau\right)  +\frac{1}{2}rn_{z}+n_{\rho}z\\
\dfrac{dn_{z}}{d\tau}=n_{\varphi}\alpha\cos\left(  \varphi-\Omega\tau\right)
-\dfrac{1}{2}rn_{\varphi}+n_{\rho}\alpha\sin\left(  \varphi-\Omega\tau\right)
.
\end{array}
\label{d8}%
\end{equation}
In this form, one is left with only \emph{three} parameters

\begin{itemize}
\item $\alpha$-the normalized strength of the rotating field,

\item $\Omega$-the normalized angular speed of the rotating field, and

\item $g$-the normalized free-fall acceleration.
\end{itemize}

\subsection{The stationary solutions and their stability.\label{sec2.2}}

We seek a solution in which the particle moves synchronously with the field at
a constant radius and height. Setting
\[
\varphi=\varphi_{0}+\Omega\tau\text{ \ ; \ }r(\tau)=r_{0}\text{ \ ; \ }%
z(\tau)=z_{0}%
\]
in Eqs.(\ref{d8}) we find \emph{two} possible solutions, given by%

\begin{equation}%
\begin{array}
[c]{c}%
n_{\rho}=-2\Omega^{2}r_{0}\\
n_{\varphi}=0\\
n_{z}=-g\\
\varphi_{0}=90^{0}\pm90^{0}\\
z_{0}=\Omega-\dfrac{g}{2\Omega^{2}r_{0}}\left(  \frac{1}{2}r_{0}\pm
\alpha\right)  .
\end{array}
\label{d9}%
\end{equation}
The value of $r_{0}$ is determined by the condition that $\left|
\mathbf{\hat{n}}\right|  =\sqrt{n_{\rho}^{2}+n_{\varphi}^{2}+n_{z}^{2}}=1$,
giving%
\begin{equation}
r_{o}=\frac{\sqrt{1-g^{2}}}{2\Omega^{2}}. \label{d10}%
\end{equation}
It can be easily shown that the stationary solution corresponding to
$\varphi_{0}=180^{0}$ has its magnetic moment \emph{antiparallel} to the
direction of the local magnetic field, whereas for the $\varphi_{0}=0^{0}$
solution it is \emph{parallel} to the direction of the field. From
Eq.(\ref{d10}) we also conclude that for a stationary solution to exist,
$\left|  g\right|  $ must be smaller than $1$. This is simply a consequence of
the fact that, in our model, the magnetic field cannot apply a force greater
than $\mu H^{\prime}$ in the $+z$-direction. When the weight of the particle
$mG$ is greater than $\mu H^{\prime}$, the magnetic force cannot balance the
weight of the particle. The latter then accelerates in the axial direction,
and no stationary solution exists. Note also that the particle is located
\emph{above} the origin with its spin pointing inward, even in the
\emph{absence} of gravity \cite{gravit}. This is necessary in order to have a
$z$-component of the field. The latter exerts torque on the spin that causes
it to rotate synchronously with the field.

To check the stability of the solutions found, we add first-order
perturbations. We set
\begin{equation}%
\begin{array}
[c]{c}%
r(\tau)=r_{0}+\delta r\\
\varphi=\Omega\tau+90^{0}\pm90^{0}+\delta\varphi\\
z\left(  \tau\right)  =\left[  \Omega-\dfrac{g}{2\Omega^{2}r_{0}}\left(
\frac{1}{2}r_{0}\pm\alpha\right)  \right]  +\delta z\\
n_{\rho}=-2\Omega^{2}r_{0}+\delta n_{\rho}\\
n_{\varphi}=0+\delta n_{\varphi}\\
n_{z}=-g+\delta n_{z},
\end{array}
\label{d11}%
\end{equation}
substitute these into Eqs.(\ref{d8}), and retain only first-order terms. We
find that the resulting equations for the perturbations are%
\begin{equation}%
\begin{array}
[c]{c}%
\dfrac{d^{2}\delta r}{d\tau^{2}}-2\Omega r_{0}\dfrac{d\delta\varphi}{d\tau
}-\delta r\Omega^{2}=-\dfrac{g}{4\Omega^{2}r_{0}}\delta n_{z}\\
2\Omega\dfrac{d\delta r}{d\tau}+r_{0}\dfrac{d^{2}\delta\varphi}{d\tau^{2}%
}=\dfrac{1}{2}\delta n_{\varphi}\\
\dfrac{d^{2}\delta z}{d\tau^{2}}=-\delta n_{z}\\
\dfrac{d\delta n_{\rho}}{d\tau}=\dfrac{g}{2\Omega^{2}r_{0}}\left(  \dfrac
{1}{2}r_{0}\pm\alpha\right)  \delta n_{\varphi}\mp\alpha g\delta\varphi\\
\dfrac{d\delta n_{\varphi}}{d\tau}-2\Omega^{2}r_{0}\dfrac{d\delta\varphi
}{d\tau}=-\dfrac{1}{2}g\delta r-2\Omega^{2}r_{0}\delta z+\dfrac{\left(
\frac{1}{2}r_{0}\pm\alpha\right)  }{4\Omega^{4}r_{0}^{2}}\delta n_{z}.
\end{array}
\label{d12}%
\end{equation}
In deriving these equations we used the constraint
\begin{equation}
\mathbf{\hat{n}\cdot}\delta\mathbf{\hat{n}=}2\Omega^{2}r_{0}\delta n_{\rho
}+g\delta n_{z}=0, \label{d13}%
\end{equation}
since $\mathbf{\hat{n}}$ is, by definition, a unit vector. We have also
discarded in Eqs.(\ref{d12}) the equation corresponding to the \emph{last}
equation in Eqs.(\ref{d8}) as it turned out to be identical to the fourth.

Looking back at Eqs.(\ref{d12}) we note that the two possible solutions differ
by the\emph{ sign }of\emph{ }$\alpha$ (reversal of the \emph{direction }of the
rotating field). It is therefore suffices to concentrate on the solution with
$\varphi_{0}=180^{0}$ for example, and to study both \emph{positive} and
\emph{negative} values of $\alpha$. This is what we do in the following.

To look for oscillatory (stable) solutions for Eqs.(\ref{d12}) we set
\begin{align*}
\delta r  &  =\left(  \delta r\right)  _{0}e^{-i\omega t}\text{ }\\
\text{ }\delta\varphi &  =\left(  \delta\varphi\right)  _{0}e^{-i\omega
t}\text{ }\\
\text{ }\delta z  &  =\left(  \delta z\right)  _{0}e^{-i\omega t}\\
\delta n_{z}  &  =\left(  \delta n_{z}\right)  _{0}e^{-i\omega t}\\
\text{\ \ \ \ \ }\delta n_{\varphi}  &  =\left(  \delta n_{\varphi}\right)
_{0}e^{-i\omega t}%
\end{align*}
inside Eqs.(\ref{d12}) and get
\begin{equation}
\underset{\mathbf{A}}{\underbrace{\left(
\begin{array}
[c]{ccccc}%
\omega^{2}+\Omega^{2} & -2i\omega\Omega r_{0} & 0 & -\dfrac{g}{4\Omega
^{2}r_{0}} & 0\\
-2i\omega\Omega & -r_{0}\omega^{2} & 0 & 0 & -\frac{1}{2}\\
0 & 0 & \omega^{2} & -1 & 0\\
0 & -\alpha g & 0 & -\dfrac{i\omega g}{2\Omega^{2}r_{0}} & \frac{g\left(
\alpha+\frac{1}{2}r_{0}\right)  }{2\Omega^{2}r_{0}}\\
\frac{1}{2}g & 2i\omega\Omega^{2}r_{0} & 2\Omega^{2}r_{0} & -\frac{\left(
\alpha+\frac{1}{2}r_{0}\right)  }{4\Omega^{4}r_{0}^{2}} & -i\omega
\end{array}
\right)  }}\cdot\left(
\begin{array}
[c]{c}%
\left(  \delta r\right)  _{0}\\
\left(  \delta\varphi\right)  _{0}\\
\left(  \delta z\right)  _{0}\\
\left(  \delta n_{z}\right)  _{0}\\
\left(  \delta n_{\varphi}\right)  _{0}%
\end{array}
\right)  =\left(
\begin{array}
[c]{c}%
0\\
0\\
0\\
0\\
0
\end{array}
\right)  .
\end{equation}
$\allowbreak$This equation has non-trivial solutions whenever the determinant
of the matrix $\mathbf{A}$ vanishes. Thus, the equation
\begin{equation}
16r_{0}\Omega^{4}\det\mathbf{A=}A_{4}\omega^{8}+A_{3}\omega^{6}+A_{2}%
\omega^{4}+A_{1}\omega^{2}+A_{0}=0, \label{d15}%
\end{equation}
where%
\begin{align}
A_{0}  &  =-32\Omega^{10}\alpha r_{0}^{3}\label{d16}\\
A_{1}  &  =+48\Omega^{8}r_{0}^{4}+4\Omega^{4}\alpha^{2}+64\alpha\Omega
^{8}r_{0}^{3}\nonumber\\
&  +2\Omega^{4}\alpha r_{0}-2g^{2}r_{0}\Omega^{4}\alpha\nonumber\\
A_{2}  &  =-16\Omega^{6}r_{0}^{4}-8\alpha^{2}\Omega^{2}-3\Omega^{2}r_{0}%
^{2}\nonumber\\
&  -16\Omega^{8}r_{0}^{2}-32\Omega^{6}r_{0}^{3}\alpha-10\alpha\Omega^{2}%
r_{0}\nonumber\\
&  -16\Omega^{5}gr_{0}^{2}-2g^{2}r_{0}\Omega^{2}\alpha-g^{2}r_{0}^{2}%
\Omega^{2}\nonumber\\
A_{3}  &  =4\alpha^{2}+r_{0}^{2}+4\alpha r_{0}+32\Omega^{6}r_{0}%
^{2}\nonumber\\
A_{4}  &  =-16\Omega^{4}r_{0}^{2},\nonumber
\end{align}
determines the eigenfrequencies $\omega$ of the various possible modes.

At this point we pause and prove that the solution with $\varphi_{0}=0^{0}$ is
\emph{not} stable for \emph{any} combination of $\alpha$, $\Omega$ and $g$: We
consider the fourth order polynomial Eq.(\ref{d15}) as a polynomial in
$x=\omega^{2}$. We note that when $x=0$, the polynomial takes on the value
$A_{0}=-32\Omega^{10}\alpha r_{0}^{3}$ which is \emph{positive} when $\alpha$
is negative. When $x\rightarrow-\infty$ on the other hand, it takes on a value
which is asymptotic to $A_{4}x^{4}=-16\Omega^{4}r_{0}^{2}x^{4}$ which, for
sufficiently large (negative) $x$, gives a \emph{negative} value. Thus, at
least one root $x=\omega^{2}$ of the polynomial must be \emph{negative},
corresponding to two\emph{ }purely \emph{imaginary} frequencies with opposite
signs, and hence to an \emph{unstable} solution.

\section{Analysis of the TOP trap of Ref.\cite{cornell}.\label{sec3}}

We now put Eq.(\ref{d15}) to use by calculating the mode frequencies of the
TOP trap described in Ref.\cite{cornell}. The parameters in this trap were:
$H=10^{-3}Tesla$, $G\simeq10m/\sec^{2}$, $H^{\prime}=2.4Tesla/meter$,
$m\simeq(A_{Rb}/A_{H})m_{proton}=1.416\cdot10^{-25}[Kg]$ (where $A_{Rb}$ is
the atomic mass of Rubidium, $A_{H}$ is the atomic mass of Hydrogen and
$m_{p}$ is the mass of proton), $f_{rotation}=7.5KHz$, $\mu=\mu_{B}%
/2=4.6\cdot10^{-24}Joule/Tesla$ and $S=\hbar=1\cdot10^{-34}Joule\cdot sec$.
From these parameters we find that $\Omega_{0}=2.049\cdot10^{4}rad/sec$,
$R_{0}=1.856\cdot10^{-7}meter$, $\alpha=2.245\cdot10^{3}$, $g=0.1283$,
$\Omega=2.3$, $r_{0}=0.0937$, $z_{0}=-288$, $n_{\rho}=-0.99$, $n_{z}=-0.1283$
and $n_{\varphi}=0$. These correspond to a particle orbiting at a radius of
$\rho_{0}=17.4nm$ and height $z_{0}=53.5\mu m$ above the origin. The magnetic
field at that point is about $-\left(  \mathbf{\hat{\rho}-}0.128\mathbf{\hat
{z}}\right)  \cdot10^{-3}Tesla$, and is therefore dominated by the strength of
the rotating field.

Solving for the four roots $\omega$ of the characteristic polynomial
Eq.(\ref{d15}) and multiplying the $\omega$'s by $\Omega_{0}/2\pi$ gives the
frequencies: $7.38MHz$, $7.524441522KHz$, $7.475558498KHz$ and $67.99Hz$. The
highest frequency $7.38MHz$ correspond to the precessional mode of motion. Its
frequency is very close to the value $\mu H_{s}/2\pi S$ where $H_{s}$ is the
strength of the local magnetic field at the stationary point. As was
calculated above, $H_{s}$ is roughly equal to the strength of the rotating
field $H$. The two middle frequencies $\sim7.5KHz$ turn out to be very close
to the frequency of rotation of the field $7.5KHz$. The reason that we keep so
many digits in these frequencies will be clarified shortly. Calculation of the
eigenvectors for these two modes yields $\left(  \delta r\right)  _{0}/\left(
\delta z\right)  _{0}\sim1.9\cdot10^{5}\gg1$ for both, which shows that their
coupling to the axial coordinate is small, indicating that these are the two
\emph{lateral} vibrational modes. Recall that the frequencies that are found
here correspond to the \emph{rotating }frame. To find the frequencies in the
\emph{laboratory} frame we explicitly calculate the time dependence of the
lateral coordinates: The $x$ coordinate for example, is given by $x\left(
\tau\right)  =\left(  r_{0}+\delta r\right)  \cos\left[  180^{0}+\Omega
\tau+\delta\varphi\right]  $ where $\delta r=\left(  \delta r\right)  _{0}%
\cos\left(  \omega\tau\right)  $, $\delta\varphi=\left(  \delta\varphi\right)
_{0}\cos\left(  \omega\tau+\phi_{0}\right)  $ and $\phi_{0}$ is a phase which
is determined by the eigenvectors. Expansion of $x\left(  \tau\right)  $ to
first order in the perturbations shows that it contains a term proportional to
the product $\cos\left[  180^{0}+\Omega\tau+\delta\varphi\right]  \cos\left(
\omega\tau+\phi_{0}\right)  $ which contains a term with \emph{slow }frequency
$\omega-\Omega$ and another term with \emph{fast }frequency $\omega+\Omega$
with the \emph{same} amplitude. We thus find that, in the laboratory frame,
the frequencies of the \emph{axial} vibrational modes are given by subtracting
and adding the rotation frequency from the middle frequencies found by using
Eq.(\ref{d15}).This gives the two \emph{slow} lateral frequencies
$+24.441522Hz$ and $-24.441502Hz$, and the \emph{fast }lateral frequencies
$15.0244KHz$ and $14.9756KHz$. The eigenvector for the lowest frequency
$67.99Hz$ satisfy $\left(  \delta r\right)  _{0}/\left(  \delta z\right)
_{0}\sim5\cdot10^{-6}\ll1$, which is a clear indication that this is the
\emph{axial} vibrational mode. In this case, the frequency of vibration is the
same both for the rotating frame and the laboratory frame so no
subtraction/addition of $\Omega$ is needed.

Table I compares our results to the measured and calculated results reported
in \cite{cornell}. We conclude that the Time-averaged Orbiting Potential (TOP)
approximation is indeed a \emph{very }good approximation to the exact result
obtained here. Note however, that the two \emph{fast} lateral frequencies
$\omega+\Omega$ that our analysis yield, were not reported in
Ref.\cite{cornell}, possibly because they were not looked for.

Note also that in the time-averaged orbiting potential approximation, the
resulting potential is necessarily isotropic in the lateral plane. Hence, in
this approximation there is only \emph{one} lateral vibrational frequency. The
exact analysis presented here, on the other hand, gives naturally \emph{two
}frequencies, corresponding to the two \emph{slow }lateral vibrational modes,
and \emph{two} more frequencies, corresponding to the \emph{fast} lateral
modes. For the TOP trap parameters given above we find the \emph{slow}
frequencies to be \emph{very} nearly equal, up to the seventh significant
digit. The splitting is a consequence of the fact that the clock-wise lateral
vibrational mode is not equivalent to the counter clock-wise lateral
vibrational mode since in both cases the spin precesses in the \emph{same}
direction. This splitting may be \emph{large} for other choice for the parameters.

\section{Approximate expressions for the mode frequencies.\label{sec4}}

For the numerical example of the previous section we have found that $g\ll1$,
$\alpha\gg1$ and $\Omega\sim1$. These relations hold also for more recent TOP
traps \cite{top1,top2}, and it is therefore natural to find approximate
expressions for the mode frequencies under these limits. We expect that for
these values, the four frequencies will be as follows: A very high frequency
$\omega_{p}$ corresponding to the precession, two very close middle
frequencies $\omega_{xy}$ which are almost equal to $\Omega$ and correspond to
the lateral vibrations, and a very low frequency $\omega_{z}$ corresponding to
the axial vibrations. The large dynamic range spanned by these frequencies
suggests that there is a certain relation between the coefficients of the
polynomial in the secular equation Eq.(\ref{d15}). We exploit this relation in
order to find approximate expressions for the mode frequencies: We first
construct a fourth-order polynomial in $\omega^{2},$ whose roots $\omega
_{p}^{2},$ $\omega_{xy}^{2}$ and $\omega_{z}^{2}$ satisfy the relations
$\omega_{p}^{2}\gg\omega_{xy}^{2}\gg\omega_{z}^{2}$, expand it in powers of
$\omega^{2}$, and keep only \emph{dominant} terms in the coefficients. This gives%

\begin{align}
&  \left(  \omega^{2}-\omega_{p}^{2}\right)  \left(  \omega^{2}-\omega
_{xy}^{2}\right)  ^{2}\left(  \omega^{2}-\omega_{z}^{2}\right) \label{d17}\\
&  \simeq\omega^{8}-\left(  \omega_{p}^{2}\right)  \omega^{6}+\left(
2\omega_{p}^{2}\omega_{xy}^{2}\right)  \omega^{4}-\left(  \omega_{p}^{2}%
\omega_{xy}^{4}\right)  \omega^{2}+\omega_{p}^{2}\omega_{xy}^{4}\omega_{z}%
^{2}.\nonumber
\end{align}
Comparing this with Eq.(\ref{d15}) shows that
\begin{align}
\omega_{p}^{2}  &  \simeq\lim_{g=0,\alpha\gg1}\left(  -\dfrac{A_{3}}{A_{4}%
}\right)  =\alpha^{2}\label{d18}\\
\omega_{xy}^{2}  &  \simeq\lim_{g=0,\alpha\gg1}\left(  -\dfrac{A_{2}}{2A_{3}%
}\right)  =\Omega^{2}\nonumber\\
\omega_{z}^{2}  &  \simeq\lim_{g=0,\alpha\gg1}\left(  -\dfrac{A_{0}}{A_{1}%
}\right)  =\dfrac{1}{\alpha}.\nonumber
\end{align}
Clearly, this approximation is not sufficient to determine the difference of
the frequencies of the \emph{lateral} vibrational modes. To furnish
\emph{these} differences, we substitute $\omega^{2}\rightarrow\Omega^{2}+d$
into Eq.(\ref{d15}) and expand to \emph{second }order in $d$. The result is
\begin{equation}
d^{2}\sum_{n=0}^{4}\dfrac{n(n-1)}{2}A_{n}\left(  \Omega^{2}\right)
^{n-1}+d\sum_{n=0}^{4}nA_{n}\left(  \Omega^{2}\right)  ^{n-1}+\sum_{n=0}%
^{4}A_{n}\left(  \Omega^{2}\right)  ^{n}=0.
\end{equation}
Setting $g=0$ and solving for $d$ gives%
\begin{equation}
d=\pm\dfrac{\Omega}{\sqrt{2\alpha}}+\mathcal{O}\left(  \alpha^{-1}\right)  ,
\label{d19}%
\end{equation}
corresponding to the \emph{slow }laboratory-frame frequencies
\begin{equation}
\left(  \omega_{xy}\right)  _{lab}=\sqrt{\Omega^{2}+d}-\Omega=\pm\dfrac
{1}{\sqrt{8\alpha}}+\mathcal{O}\left(  \alpha^{-3/2}\right)  \text{.}
\label{d20}%
\end{equation}
$\allowbreak$In particular we find that
\begin{equation}
\dfrac{\omega_{z}}{\left(  \omega_{xy}\right)  _{lab}}=\pm\sqrt{8}%
+\mathcal{O}\left(  \alpha^{-1/2}\right)  , \label{eq34}%
\end{equation}
which shows that our analysis reduces to the results reported in
Ref.\cite{cornell}.

In non-normalized units the mode frequencies in the laboratory-frame are given
by%
\begin{align*}
\omega_{prec.}  &  \simeq\dfrac{\mu H}{S}\\
\omega_{z}\Omega_{0}  &  \simeq\pm\sqrt{\dfrac{\mu H^{\prime2}}{8mH}}\\
\left(  \omega_{xy}\right)  _{lab}\Omega_{0}  &  \simeq\sqrt{\dfrac{\mu
H^{\prime2}}{mH}}.
\end{align*}

\section{The stability region for $g=0$.\label{sec5}}

Having analyzed the TOP trap for the limit $\alpha\gg1$, we turn our attention
to study the region $\alpha$, $\Omega\sim1$, which has not yet been exploited
experimentally. To keep matters simple, we specify to the gravitation-free
case $g=0$. We use Eq.(\ref{d15}) to scan the $\alpha$-$\Omega$ plane in the
search of points corresponding to stable solutions. The result is shown in
Fig.(\ref{fig1}).

This figure shows the \emph{boundary} lines between points corresponding to
stable regions and unstable regions. The fact that these lines consist of
different segments, indicate that two different segments of a given line
correspond to \emph{different} modes that become unstable. For example, going
upward along the $\alpha=2.5$ line, we find that the two \emph{slowest} modes
coalesce and become unstable at $\Omega\simeq0.73$. As $\Omega$ is increased,
these modes become stable again at $\Omega\simeq0.91$. When $\Omega$ is
further increased, the two \emph{fastest} modes become unstable at
$\Omega\simeq1.72$. Note that we have found a small stability region in the
range $\Omega\simeq1.2$-$2$ and $\alpha\simeq0$-$0.05$, which we did not
investigate in detail.

As $A_{0}\propto\Omega^{10}\alpha$ we conclude that points, both along the
$\Omega=0$ line and along the $\alpha=0$ line, have one mode with vanishing
frequency, corresponding to a soft mode. In addition, all points in the
$\alpha<0$ half-plane correspond to unstable solutions, as was proved earlier.
Note also that the coefficients in the secular equation really depend on
$\Omega^{2}$ and not just on $\Omega$. Hence, the continuation of the
stability diagram to the $\Omega<0$ half-plane is simply a mirror reflection
of the $\Omega>0$ half-plane with respect to the $\Omega=0$ line.

The stability diagram shows that the TOP trap is much more tolerant that what
one would have expected. There are many points near the corner of the first
quadrant of the $\alpha$-$\Omega$ plane that may be used experimentally. Note
however that as $r_{0}\sim1/\Omega^{2}$, the use of too low a value for
$\Omega$ results in a large radius. This radius may fall outside the region in
which our linear approximation to the spatial dependence of the field holds.
The lower bound on $\Omega$ is therefore determined by the \emph{second}
derivative of the field. In addition, quantum-mechanical considerations set a
lower bound on $\alpha$ as well. It can be shown that the extent of the
wavefunction of the particle $\Delta x_{\text{quantum}}\sim\sqrt{\hbar
/m\omega}$, and the extent of the field $\Delta x_{\text{field}}\sim
H/H^{\prime}$ scale as $\Delta x_{\text{quantum}}/\Delta x_{\text{field}}%
\sim\alpha^{-3/4}$. Thus, in order the keep the extent of the wavefunction
much smaller than the extent of the field, $\alpha$ too must be kept large enough.

\section{Connection with the adiabatic approximation.}

It is instructive to study the same problem in the limit where the system is
\emph{extremely} adiabatic. For simplicity we specify to the case $G=0$. In
this approximation, the direction of the spin $\mathbf{\hat{n}}$ is
\emph{locked} to the direction of the \emph{local} magnetic field, so that%
\begin{equation}
\mathbf{\hat{n}\simeq}\dfrac{\mathbf{H-}\dfrac{S}{\mu}\Omega_{r}%
\mathbf{\hat{z}}}{\left|  \mathbf{H-}\dfrac{S}{\mu}\Omega_{r}\mathbf{\hat{z}%
}\right|  }\text{.} \label{ad1}%
\end{equation}
Substitution of Eq.(\ref{ad1}) into Eq.(\ref{d6.0}) and discarding the
equation for the spin Eq.(\ref{d6.2}) gives%
\begin{equation}
m\left[  \dfrac{d^{2}\mathbf{r}}{dt^{2}}+2\Omega_{r}\mathbf{\hat{z}}%
\times\dfrac{d\mathbf{r}}{dt}+\Omega_{r}^{2}\mathbf{\hat{z}\times}\left(
\mathbf{\hat{z}}\times\mathbf{r}\right)  \right]  =-\mu\mathbf{\nabla}\left|
\mathbf{H-}\dfrac{S}{\mu}\Omega_{r}\mathbf{\hat{z}}\right|  , \label{ad2}%
\end{equation}
with $\mathbf{H}$ given by Eq.(\ref{d7}).

It is important to note that this approximation is \emph{different} from the
time-averaged orbiting potential (TOP) approximation. In the latter, one
constructs a time-dependent potential $V_{TOP}(\mathbf{r},t)\propto\left|
\mathbf{H}\left(  \mathbf{r},t\right)  \right|  $, then averages
$V_{TOP}(\mathbf{r},t)$ over time to get a time-\emph{independent} potential
$V_{TOP}^{0}\left(  \mathbf{r}\right)  $. Here, we work in the \emph{rotating}
frame, in which $\mathbf{H}\left(  \mathbf{r}\right)  $ is
time-\emph{independent}, and construct a time-independent potential
\[
V_{AD}(\mathbf{r})\propto\left|  \mathbf{H-}\dfrac{S}{\mu}\Omega
_{r}\mathbf{\hat{z}}\right|  .
\]

Normalizing the Cartesian components of the position vector $\mathbf{r}%
=x\mathbf{\hat{x}}+y\mathbf{\hat{y}}+z\mathbf{\hat{z}}$ to $R_{0}$, we find
that a stationary solution to Eq.(\ref{ad2}) is given by%
\begin{equation}%
\begin{array}
[c]{c}%
x_{0}=-\dfrac{1}{2\Omega^{2}}\\
y_{0}=0\\
z_{0}=\Omega.
\end{array}
\label{ad3}%
\end{equation}
This result agrees with Eqs.(\ref{d9}) and (\ref{d10}) for the case $g=0$. The
other possible solution, in which $x_{0}$ is \emph{positive,} is discarded
because it is not stable.

Substituting%
\begin{equation}%
\begin{array}
[c]{c}%
x=x_{0}+\delta x\\
y=y_{0}+\delta y\\
z=z_{0}+\delta z
\end{array}
\label{ad4}%
\end{equation}
into the normalized form of Eq.(\ref{ad2}), and expanding to first order in
the perturbations gives%
\begin{equation}%
\begin{array}
[c]{c}%
\dfrac{d^{2}\delta x}{dt^{2}}-2\Omega\dfrac{d\delta y}{dt}-\Omega^{2}\delta
x=0\\
\dfrac{d^{2}\delta y}{dt^{2}}+2\Omega\dfrac{d\delta x}{dt}-\Omega^{2}\delta
y=-\dfrac{\Omega^{2}}{4\Omega^{2}\alpha+1}\delta y\\
\dfrac{d^{2}\delta z}{dt^{2}}=-\dfrac{4\Omega^{2}}{4\Omega^{2}\alpha+1}\delta
z.
\end{array}
\label{ad5}%
\end{equation}
We therefore find that the axial translational degree of freedom is decoupled
from the rest, with a frequency%
\[
\omega_{z}=\dfrac{2\Omega}{\sqrt{4\Omega^{2}\alpha+1}}\simeq\dfrac{1}%
{\sqrt{\alpha}}+O\left(  \alpha^{-3/2}\right)  \text{ },
\]
which agrees with Eq.(\ref{d18}). For the lateral translational degrees of
freedom we have%
\[
\underset{\mathbf{M}}{\underbrace{\left(
\begin{array}
[c]{cc}%
-\omega^{2}-\Omega^{2} & 2i\omega\Omega\\
-2i\omega\Omega & -\omega^{2}-\Omega^{2}+\dfrac{\Omega^{2}}{4\Omega^{2}%
\alpha+1}%
\end{array}
\right)  }}\cdot\left(
\begin{array}
[c]{c}%
\delta x\\
\delta y
\end{array}
\right)  =\left(
\begin{array}
[c]{c}%
0\\
0
\end{array}
\right)  ,
\]
for which a non-trivial solution exists whenever%
\[
\left(  4\Omega^{2}\alpha+1\right)  \det\mathbf{M}=\left(  4\Omega^{2}%
\alpha+1\right)  \omega^{4}-\left(  8\Omega^{4}\alpha+3\Omega^{2}\right)
\omega^{2}+4\Omega^{6}\alpha=0.
\]
This equation furnishes the frequencies%
\[
\omega_{xy}^{2}=\dfrac{8\Omega^{4}\alpha+3\Omega^{2}\pm\Omega^{2}%
\sqrt{32\Omega^{2}\alpha+9}}{2\left(  4\Omega^{2}\alpha+1\right)  },
\]
corresponding to the laboratory-frame frequencies%
\[
\left(  \omega_{xy}\right)  _{lab}=\omega_{xy}-\Omega=\pm\frac{1}%
{\sqrt{8\alpha}}+\mathcal{O}\left(  \alpha^{-3/2}\right)  .
\]
$\allowbreak$This result also agrees with Eq.(\ref{d20}).

Though in this model the spin is locked to the direction of the field, we can
nevertheless define a precessional frequency $\omega_{p}$ by calculating the
field \emph{at the stationary point} $H_{s}$ and define $\omega_{p}\Omega
_{0}\equiv\mu H_{s}/S$. This gives%
\[
\omega_{p}=\alpha+\dfrac{1}{4\Omega^{2}},
\]
which for $\alpha\Omega^{2}\gg1$ coincides with Eq.(\ref{d18}).

Note however, that the adiabatic approximation presented here holds whenever
the precession speed $\omega_{p}$ is \emph{large} compared to the vibrational
frequencies. In addition, $\omega_{p}$ should also be large compared to the
rotation frequency $\Omega$. The extreme case $\omega_{p}=\Omega$ defines a
line in the $\alpha$-$\Omega$ plane which for large $\alpha$ approaches
asymptotically to the line $\alpha=\Omega$.

\section{Discussion\label{sec6}}

We have shown that our exact analytic results reduce to the formulae derived
via the time-averaged orbiting potential approximation, in the case where
$\alpha$ is large. In addition, the stability diagram that we found, suggests
that the TOP trap is very flexible for the experimentalist in terms of the
allowed parameters. We have also shown that under the adiabatic approximation,
where the direction of the spin $\mathbf{\hat{n}}$ is locked to the direction
of the field in the \emph{rotating} frame, we recover, for large magnetic
field $\alpha$, the exact mode frequencies.

It is interesting to note that Eqs.(\ref{d6.0}) and (\ref{d6.2}) pave the way
for a \emph{quantum-mechanical} treatment of the same problem in the comoving
frame. When gravity is neglected, the Hamiltonian for this system is given by%
\[
\hat{H}=\dfrac{\left(  \mathbf{\hat{P}}-\mathbf{A}\right)  ^{2}}{2m}%
+\mu\mathbf{\sigma}^{S}\mathbf{\cdot}\left(  \mathbf{H-}\dfrac{S}{\mu}%
\Omega_{r}\mathbf{\hat{z}}\right)  -\dfrac{1}{2}m\Omega_{r}^{2}\left(
x^{2}+y^{2}\right)  ,
\]
where $\mathbf{A}$ is a vector potential field satisfying
\[
\mathbf{\nabla\times A}\propto\Omega_{r}\mathbf{\hat{z}}\text{,}%
\]%
\[
\mathbf{\sigma}^{S}\mathbf{=}\sigma_{x}^{S}\mathbf{\hat{x}}+\sigma_{y}%
^{S}\mathbf{\hat{y}+}\sigma_{z}^{S}\mathbf{\hat{z}}%
\]
$\ $\ is the spin $S$ (where $S$ could be either $0$ or $1/2$ or $1$ etc.)
vector of Pauli matrices, $\mathbf{\hat{P}}$ is the vector momentum operator,
$\mathbf{H-}\dfrac{S}{\mu}\mathbf{\Omega}_{r}$ is the magnetic field as seen
in the rotating frame, and $-m\Omega_{r}^{2}\left(  x^{2}+y^{2}\right)  /2$ is
the centrifugal potential. In order to diagonalize the magnetic part of the
Hamiltonian, one performs a local \emph{passive} transformation of coordinates
on the wave function, such that the spinor is expressed in a new coordinate
system whose $z$-axis coincides with the direction of the local magnetic field
$\mathbf{H-}\dfrac{S}{\mu}\mathbf{\Omega}_{r}$ at the point $\mathbf{r}$. This
rotation does not affect either the centrifugal term or $\mathbf{A}$. The
momentum $\mathbf{\hat{P}}$ however, transforms to $\mathbf{\hat{P}%
}-\mathbf{A}^{\prime}\left(  \mathbf{r,}\sigma_{x}^{S},\sigma_{y}%
^{S}\mathbf{,}\sigma_{z}^{S}\right)  $, where $\mathbf{A}^{\prime}\left(
\mathbf{r,}\sigma_{x}^{S},\sigma_{y}^{S}\mathbf{,}\sigma_{z}^{S}\right)  $
contains \emph{non-diagonal} elements as it includes the spin degree of
freedom. For typical values of parameters, the non-diagonal part may be
treated as a small perturbation, and the \emph{lifetime} of the particle in
the trap may be calculated. This technique has already been applied
successfully to a 1D toy-model \emph{time-independent }magnetic trap\cite{1d},
and to a Ioffe-like 2D trap\cite{2d}, for the case of spin $S=1/2$ particles.
As the TOP trap is used to capture \emph{Bosons}, it is more resonable to
study it for the case $S=1$. We believe that despite the complexity that the
Coriolis and centrifugal forces add to the problem, it is possible to solve
TOP trap \emph{quantum-mechanically}. The analysis of this problem is still
under study.

\section{Acknowledgment}

It is our pleasure to acknowledge with thanks Prof. H. Thomas for many helpful
discussions of the physics of the TOP, which clarified to us the subtleties of
this ingenuous scheme.

\newpage

\begin{table}[ptb]
\caption{Comparison of mode frequencies.}%
\centering
\par%
\begin{tabular}
[c]{|l|l|l|l|}\hline
Mode & Measured \cite{cornell} & TOP \cite{cornell} & Exact Anlaysis\\\hline
Prec. freq. $MHz$ & - & $\sim7$ & $7.38$\\
Axial freq. $Hz$ & $67\pm1$ & $69\pm2$ & $67.99$\\
Lateral freq. $Hz$ & $24\pm1$ & $24\pm1$ & $\left\{
\begin{array}
[c]{c}%
\pm24.44\\
15K\pm24.44
\end{array}
\right.  $\\\hline
\end{tabular}
\end{table}

\newpage

\begin{figure}[p]
\begin{center}
\end{center}
\vspace{5.3in}
\caption{Stable region for $g=0$ in the $\alpha$-$\Omega$ plane.}%
\end{figure}

\begin{figure}[ptb]
\begin{center}
\includegraphics[
height=5.3437in,
width=3.7784in
]{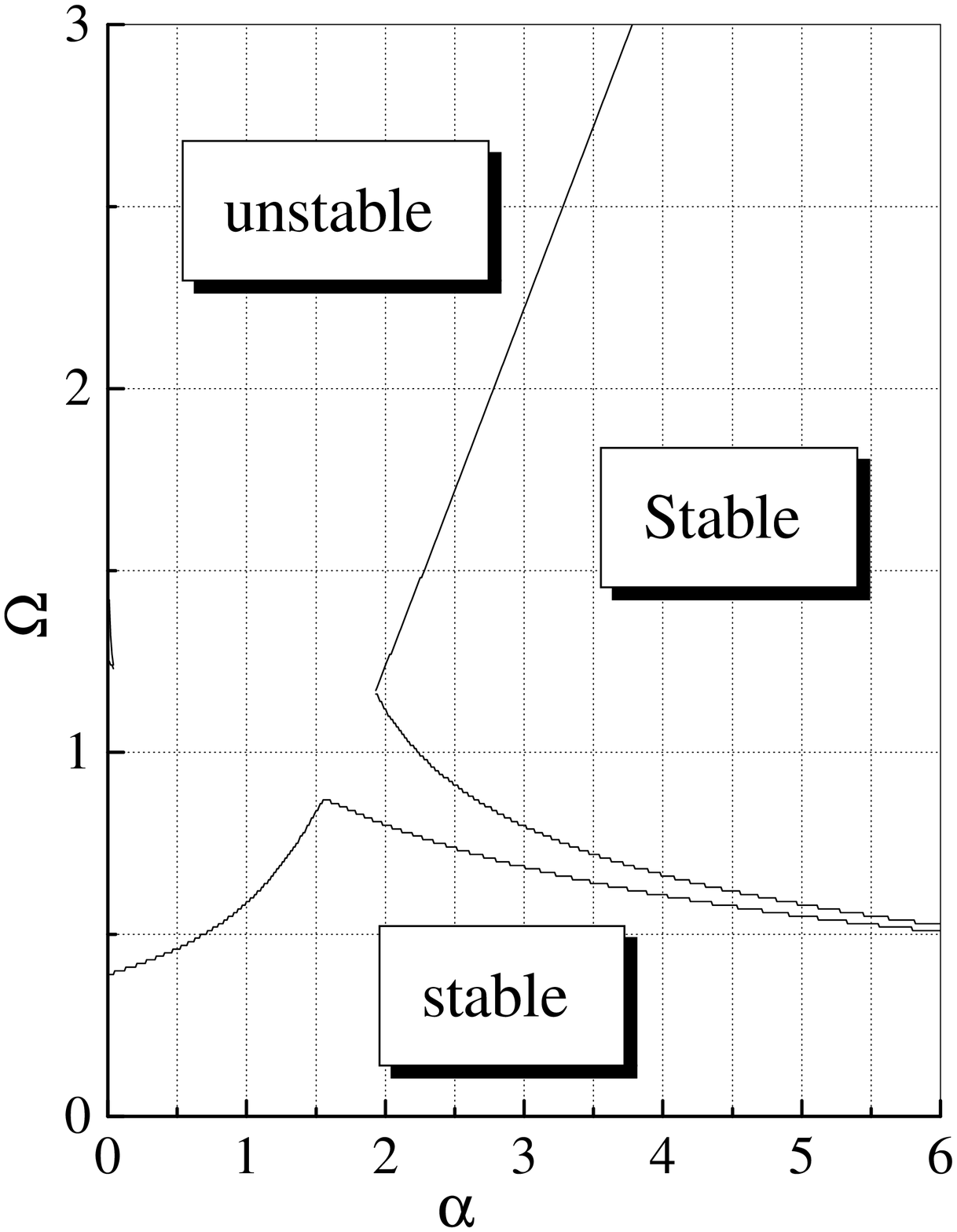}
%
%
\label{fig1}
\end{center}
\end{figure}

\end{document}